\documentclass[sigconf]{acmart}

\copyrightyear{2024}
\acmYear{2024}
\setcopyright{acmlicensed}
\acmConference[MRAC '24] {Proceedings of the 2nd International Workshop on Multimodal and Responsible Affective Computing}{November 1, 2024}{Melbourne, VIC, Australia.}
\acmBooktitle{Proceedings of the 2nd International Workshop on Multimodal and Responsible Affective Computing (MRAC '24), November 1, 2024, Melbourne, VIC, Australia}
\acmISBN{979-8-4007-1203-6/24/11}
\acmDOI{10.1145/3689092.3689414}

\usepackage{subfigure}
\usepackage{diagbox}
\usepackage{enumitem}
\usepackage{hyperref}
\usepackage{amsmath}
\usepackage{booktabs}
\usepackage{threeparttable}
\usepackage{array}
\usepackage{graphicx}
\usepackage{multirow}
\usepackage{balance}

\newcolumntype{C}[1]{>{\centering\arraybackslash}m{#1}}

\begin{document}

\title{Audio-Guided Fusion Techniques for Multimodal Emotion Analysis}



\author{Pujin Shi}
\affiliation{%
  \institution{
  State Key Laboratory of Networking and Switching Technology \\
  School of Cyberspace Security \\
  Beijing University of Posts and Telecommunications}
  \city{Beijing}
  \country{China}
}
\email{shipujin@bupt.edu.cn}

\author{Fei Gao}
\authornote{Corresponding author.}
\affiliation{%
  \institution{
    State Key Laboratory of Networking and Switching Technology \\
  School of Cyberspace Security \\
  Beijing University of Posts and Telecommunications
  }
  \city{Beijing}
  \country{China}
}
\email{gaof@bupt.edu.cn}


\renewcommand{\shortauthors}{Pujin Shi and Fei Gao}

\begin{abstract}
In this paper, we propose a solution for the semi-supervised learning track (MER-SEMI) in MER2024. First, in order to enhance the performance of the feature extractor on sentiment classification tasks,
we fine-tuned video and text feature extractors, specifically CLIP-vit-large and Baichuan-13B, using labeled data. This approach effectively preserves the original emotional information conveyed in the videos.  Second, we propose an Audio-Guided Transformer (AGT) fusion mechanism, which leverages the robustness of Hubert-large, showing superior effectiveness in fusing both inter-channel and intra-channel information. Third, To enhance the accuracy of the model, we iteratively apply self-supervised learning by using high-confidence unlabeled data as pseudo-labels. Finally, through black-box probing, we discovered an imbalanced data distribution between the training and test sets. Therefore, We adopt a prior-knowledge-based voting mechanism.  The results demonstrate the effectiveness of our strategy, ultimately earning us third place in the MER-SEMI track.
\end{abstract}



\maketitle

\section{Introduction}
Recently, emotion recognition has garnered attention due to its wide range of applications, such as affective computing\cite{dahl2007sleep}, healthcare\cite{feinberg1986facial}, human-computer interaction\cite{mauss2013poorer,ba2023measuring}, and market research\cite{torres2020emotion}. Traditional emotion recognition methods utilize physical and physiological signals\cite{kamble2023comprehensive}. These methods rely on specialized instruments, which can be time-consuming and labor-intensive to implement. Multimodal emotion recognition\cite{abdullah2021multimodal}, on the other hand, extracts emotional representations from speech, visual and text modalities. It employs fusion mechanisms to integrate multimodal features for downstream emotion classification tasks, offering a more flexible and efficient approach\cite{he2022multimodal}.In studying emotions, two primary approaches are commonly adopted: the dimensional approach\cite{mehrabian1996pleasure} and the discrete approach\cite{plutchik2013theories}. The objective of the MER-SEMI track (MER 2024\cite{lian2024mer}) is to map samples to six correct emotion labels (worried, happy, neutral, angry, surprise, and sad), which fundamentally falls under the task of discrete emotion label classification. This requires us to maximize the retention of emotional information from each modality during feature fusion and simultaneously minimizing the interference of irrelevant information between different modalities\cite{huang2022modality,Huang2023ContextBasedAM}.

\begin{figure*}[htbp]
    \centering
    \includegraphics[width=0.90\textwidth]{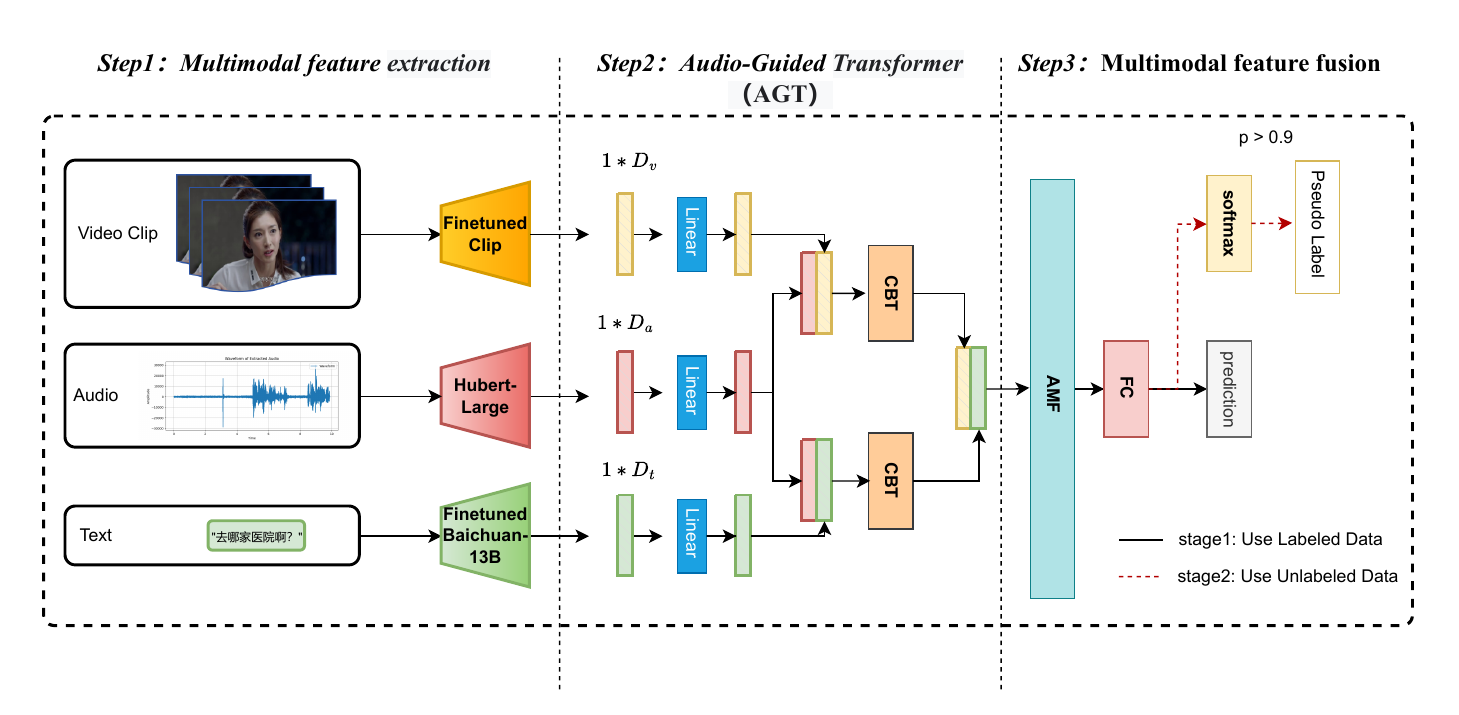}
    \label{fig:overview}
    \caption{Overview of our multimodal emotional feature fusion framework. (CBT: Context-based Transformer. AMF: Adaptive Multimodal Fusion.Stage 1: Train using labeled data; Stage 2: Generate pseudo-labels using unlabeled data and add them back to the training set for further training.)}
    \label{fig:overview}
\end{figure*}

In the MER-SEMI track, there are two primary challenges: \textbf{(1)} How to enhance the robustness ability of models using unsupervised or semi-supervised learning methods in the context of scarce labeled data;\label{sent1} \textbf{(2)} How to maximally retain the complementary emotional information across different modalities during multimodal feature fusion;\label{sent2} To address \textbf{(1)}, various methods have been proposed. For instance, Ding et al. \cite{Ding2023LearningAA} introduced the Video-Audio Transformer, which aligns visual and audio modalities and incorporates contrastive loss for semi-supervised learning, achieving significant results\cite{lian2023mer}. Similarly, Chen et al.\cite{Chen2023SemiSupervisedME} proposed a class-balanced pseudo-labeling strategy to select high-confidence pseudo-labeled samples for each category from unlabeled data. Moreover, You et al.\cite{You2021SelfsupervisedCC} designed a Temporal-Alignment Attention mechanism to align speech and textual cues in a common space and used auxiliary self-supervised tasks to ensure the consistency and coherence of unlabeled data. Additionally, Yang et al.\cite{yang2023confede} integrated inter-sample contrastive learning and intra-sample modality decomposition into a simple unified loss function, thereby simplifying the training process. To tackle \textbf{(2)}, Mittal et al.\cite{Mittal2021Affect2MMAA} proposed a learning model based on Long Short-Term Memory (LSTM) networks for emotion perception. In Chen et al.'s work \cite{Chen2023SemiSupervisedME}, large-scale unlabeled emotional video data were used to train a mask autoencoder (Expression MAE\cite{he2022masked}). Further, Wang et al.\cite{Wang2023HierarchicalAI} designed three different structures based on Attention-Guided Feature Aggregation (AFG)\cite{lian2019conversational} for deep feature fusion. Lastly, Praveen et al.\cite{Praveen2022AJC} proposed a Joint Cross-Attention model that extracts key emotional features. Overall, these methods aim to enhance the accuracy of emotion recognition models from the perspectives of pseudo-label construction and feature fusion.

Recently, researchers have noted that the contribution of the auditory modality is relatively more significant compared to visual and textual modalities\cite{Ding2023LearningAA,pastor2022cross}. 
Based on this conclusion, our main contributions are as follows:

\begin{itemize}
        \item We fine-tuned CLIP-vit-large\cite{radford2021learning} and Baichuan-13B\cite{yang2023baichuan} using labeled data, allowing the feature extractors to focus on emotional information rather than unrelated background noise. We refer to these fine-tuned feature extractors as Fine-tuned-Clip-large and Fine-tuned-Baichuan-13B.
        \item 
        We improved CAMFNET\cite{Huang2023ContextBasedAM} and proposed Audio-Guided Transformer fusion mechanism (AGT) . This method uses audio modality features as the leading input for tri-modal fusion. It leverages the generalization of Hubert-large and has been found to more effectively integrate both inter-channel and intra-channel complementary information.
        \item To leverage unlabeled data as pseudo-labels , inspired by \cite{li2023multimodal,Chen2023SemiSupervisedME,arazo2020pseudo,fang2023rethinking}, We select soft pseudo labels with a confidence higher than 0.9 for self-supervised training. Considering the imbalanced data distribution between training and test sets (Figure \ref{fig:leida}), we implemented a prior-knowledge-based voting mechanism \cite{tanha2020boosting,li2017reliable} that takes advantage of prediction distribution inconsistencies across different architectures, with special handling for imbalanced labels.
\end{itemize}
The overall framework of our model is illustrated in Figure \ref{fig:overview}.

\section{Method}
In this section, we will discuss our proposed multimodal emotion recognition system in three subsections. The selection and fine-tuning of feature extractors constitute the first part, the AGT fusion mechanism constitutes the second part, and the pseudo-label and voting mechanism constitutes the third part.


\subsection{Fine-tuning the feature extractor}
\begin{figure*}[htbp]
    \centering
    \subfigure[Fine-tuning CLIP-vit-large]{
        \includegraphics[width=0.35\textwidth]{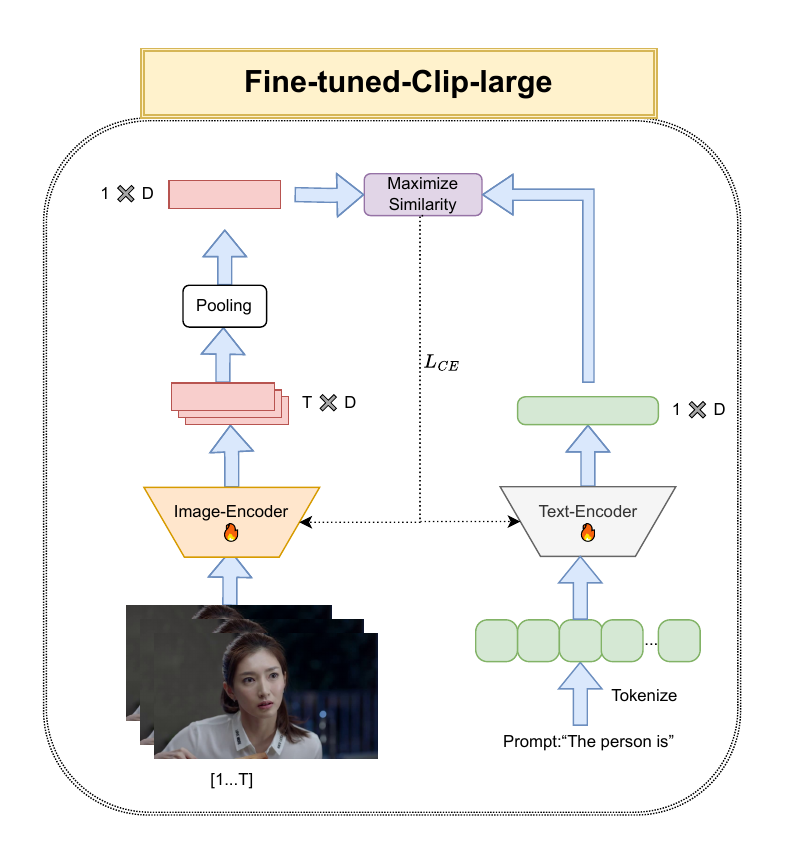}
        \label{fig:subfigA}
    }
    \hspace{0.02\textwidth}
    \subfigure[Fine-tuning Baichuan-13B]{
        \includegraphics[width=0.31\textwidth]{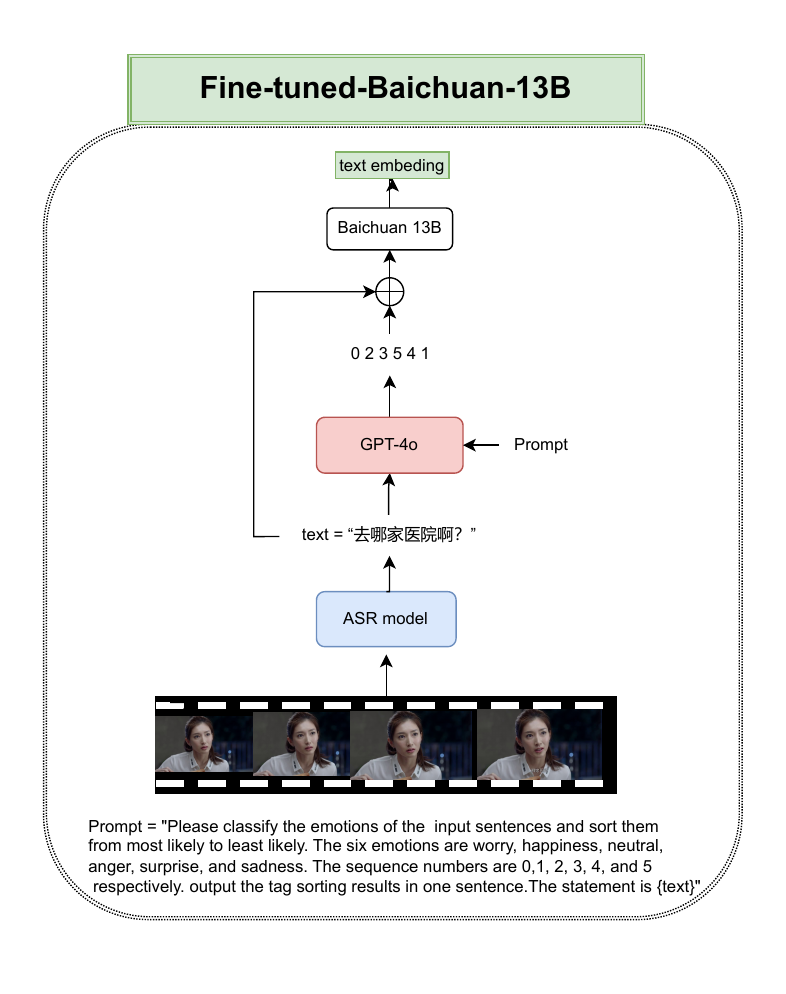}
        \label{fig:subfigB}
    }
    \caption{Fine-tuning the feature extractors for both visual and textual modalities}
    \label{fig:overall}
\end{figure*}

Feature extractors, such as Hubert-large, CLIP-vit-large, and Baichuan-13B, perform well in single-modality scenarios\cite{lian2024mer}. However, they tend to exhibit poor generalization in multi-modality settings. We hypothesize that this may be due to the fact that these feature extractors are trained on public datasets, where they may learn redundant information that is not conducive to multimodal sentiment recognition\cite{huang2022modality}.Therefore, for the audio modality, we ultimately selected Hubert-large due to its superior generalization ability. This model effectively extracts semantic information from raw audio that complements other modalities. Additionally, for the visual and textual modalities, we initially selected CLIP-large and Baichuan-13B as the feature extractors. However, since both of these large models are trained on extensive wild datasets, their use for multimodal emotion recognition introduces a substantial amount of irrelevant information (e.g., background information in videos, emotion-irrelevant information in text). Consequently, we fine-tuned these two feature extractors using the labeled data\cite{lian2024mer,lin2022frozen,Wu2022NoisyTuneAL,Xiao2024Baichuan2SumIF}. The specific fine-tuning steps are illustrated in Figure \ref{fig:overall}.

Regarding the visual modality, based on \cite{rasheed2023fine}, better alignment of visual and textual representations in video tasks provides stronger generalization capabilities. Therefore, we performed alignment operations on 5030 labeled data samples using CLIP-vit-large. As shown in Figure\ref{fig:subfigA}, given a video sample $V_i \in \mathbb{R}^{T \times H \times W \times C}$ with $T$ frames and its corresponding text label Y, the CLIP image encoder independently encodes each of the $T$ frames into a batch of images, resulting in frame-level embeddings $X_i \in \mathbb{R}^{T \times D}$. The CLIP text encoder encodes class Y, wrapped in a prompt template such as `the person is <emo-label>', to produce a text embedding $t \in \mathbb{R}^{D}$. For a batch of videos, the cosine similarity sim(.) maximizes the information between all video-level embeddings $v_i$ and their corresponding text embeddings $t_i$, fine-tuning the CLIP model using cross-entropy (CE,(\ref{eq:tempered_cross_entropy})) objective with a temperature parameter $\tau$.

\begin{equation}
\label{eq:tempered_cross_entropy}
L = - \sum_{i} \log\left(\frac{\exp(sim(v_i,t_i) / \tau)}{\sum_{j} \exp(sim(v_i,t_j) / \tau)}\right)
\end{equation}

For the textual modality,instead of fine-tuning the model, we adopted the approach of prompt tuning\cite{li2023prompt} to generate refined prompts.Specifically, we utilized GPT-4 to perform emotion recognition on text using labeled data within six predefined emotion categories (0: worry, 1: happiness, 2: neutral, 3: anger, 4: surprise, and 5: sadness), outputting the classifications in order of descending likelihood as shown in Figure \ref{fig:subfigB}. Subsequently, we concatenated the original text information with the output labels and fed this combined input into Baichuan-13B to obtain fine-tuned textual features.

\subsection{AGT Fusion Mechanism}
As illustrated in Figure \ref{fig:overview}, we extract features from each modality using the selected feature extractors. Subsequently, we use the semantic features extracted by Hubert to guide visual features and textual features independently, using the Context-Based Transformer (CBT) module \cite{Huang2023ContextBasedAM}. This module primarily extracts deep information within channels. it effectively integrates both global and local information using a multi-head self-attention mechanism, leveraging skip connections and feedforward layers to prevent overfitting and enhance the model's linear representation capabilities.
During the fusion process, we employed Adaptive Multimodal Fusion (AMF) \cite{Huang2023ContextBasedAM}, which addresses the issue of different modalities conveying disparate emotions. This method utilizes a multi-head self-attention mechanism to calculate the similarity between different modalities. AMF can determine whether a particular modality expresses an emotion distinct from the others. If a specific modality exhibits very low similarity with all other modalities, it is considered to express a different emotion, and its features are thus zeroed out to eliminate interfering information.
\begin{equation}
\label{eq:agt}
X_f = f_{AMF}{(f_{CBT}{(cat[A_i,V_i])}+f_{CBT}{(cat[A_i,T_i])})}
\end{equation}
where $X_f$ represents the fused feature, while $A_i$, $V_i$, and $T_i$ denote the audio, visual, and textual modality features of the (i)-th sample, respectively.

\subsection{Pseudo-label and Voting Mechanism}
\begin{figure}[h]
  \centering
  \includegraphics[width=\linewidth]{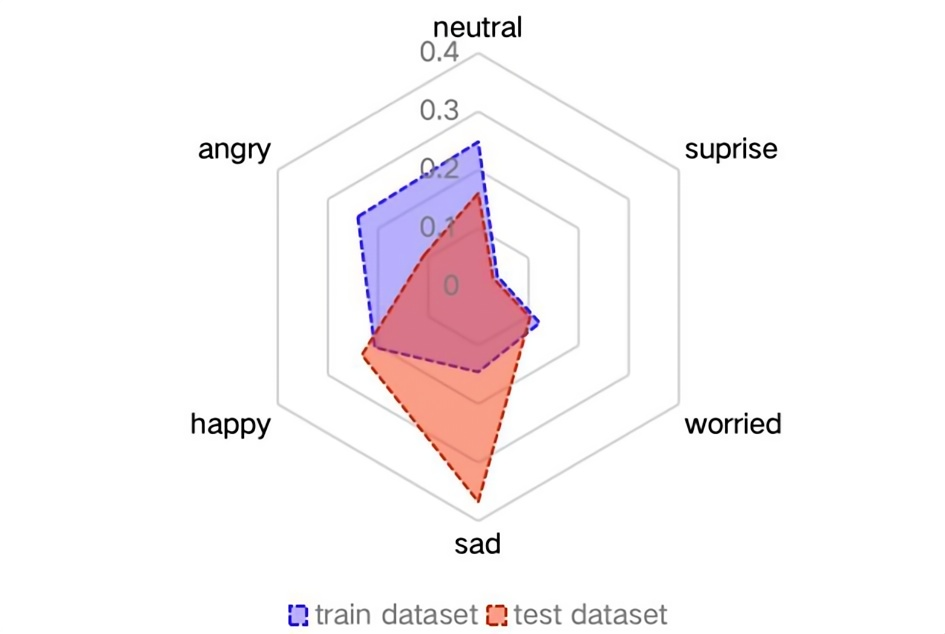}
  \caption{Distribution of the training and testing sets across the six emotional labels}
  \label{fig:leida}
\end{figure}

In order to verify if the distributions of the training and testing sets are consistent (which has implications for the necessity of targeted data augmentation), we calculated the quantity of data for each label in the training set (worried: 616, happy: 1038, neutral: 1248, sad: 730, angry: 1208, surprise: 190), and estimated the approximate distribution of the test dataset through F1-scores obtained by black-box probing (worried: 0.0326, happy: 0.0732, neutral: 0.0505, sad: 0.1157, angry: 0.03412, surprise: 0.0094). As shown in Figure\ref{fig:leida}, the statistical results indicate a significant discrepancy between the distributions of the test set and training set, particularly for the 'sad' and 'worried' labels. Consequently, low-confidence labels are prone to misclassification during pseudo-label generation (e.g., The ground truth label is 'sad', but the predicted result is 'worried).

Based on this, we first generated pseudo-labels with high confidence ( p > 0.9 ) on the test set using Hubert-large, the baseline model, and AGT, and then added the intersection of these results to the training set for iterative training. To fully leverage the generalization potential of Hubert-large, we developed a prior-knowledge-based regularized voting mechanism: when the three models make predictions for 'happy', 'neutral', 'surprise', and 'angry', we adopt a majority voting principle. However, when 'worried' and 'sad' appear in the predictions, to mitigate the impact of overfitting, we use a probabilistic approach to control the weight of predictions from each model. Specifically, 80\% of the final prediction is based on the Hubert-large results, and the remaining 20\% is from either the baseline model or AGT. This prior-knowledge-based regularized voting mechanism contributed significantly to the improvement of the F1-score.

\section{Experiments and Results}
In this section, we conduct ablation experiments and analyze the results. First, we perform comparative experiments on features from two and three modalities, using different feature extractors and fusion mechanisms. Second, we validate the effectiveness of various strategies by incorporating them into the experiments.Note that the evaluation metric for the following experiments is the F1-score.

In Table \ref{tab:perfor}, we observe that the fine-tuned feature extractors exhibit significant effectiveness when used in combination, but their performance declines when used individually. We propose a possible reason for this phenomenon: the fine-tuning alignment process may enhance the complementary capabilities between the video and text modalities, but it could also result in some modality competition between the audio and video modalities. Additionally, for the same combinations of feature extractors, the AGT fusion mechanism consistently achieves higher F1-scores on both Train\&Val and MER-SEMI datasets compared to the baseline fusion mechanism. This indicates that the AGT fusion mechanism is more effective in capturing complementary information across different modalities during cross-modal feature fusion.

Furthermore, the results in Table \ref{tab:distortion_type} indicate that the use of self-supervised pseudo-labels (which were obtained by filtering soft labels with a confidence level greater than 0.9 after applying softmax.) training methods can effectively improve the accuracy of model predictions, both in the baseline model and AGT. The proposed prior-knowledge-based voting mechanism has also demonstrated its effectiveness in ablation experiments. This method leverages the imbalanced data distribution between the training and test sets as well as the robustness of Hubert-large to regularize the prediction results with a certain probability. 

\section{Conclusion}
In this paper, we fine-tune CLIP-large and Baichuan-13B as feature extractors and propose an Audio-Guided Transformer (AGT) architecture. This fusion mechanism eliminates redundant information between modalities while capturing deeper emotional representations. Additionally, we employ self-supervised learning using pseudo-labels. Given the prior knowledge of the imbalanced distribution between training and test data, we assign higher weights to the more generalizable Hubert-large for imbalanced label predictions and use a voting mechanism to improve model prediction accuracy. Ultimately, these methods helped us achieve a score of 89.83\% in the MER-SEMI 2024 track.

\renewcommand{\arraystretch}{1.3}

\begin{table}[tp]
    \centering
    \fontsize{9}{10}\selectfont
    \begin{threeparttable}
        \caption{Performance of the baseline model and AGT across various feature combinations used as input. Specifically, 
        \normalsize{\textcircled{\scriptsize{0}}},
        \normalsize{\textcircled{\scriptsize{1}}},
        \normalsize{\textcircled{\scriptsize{2}}},
        \normalsize{\textcircled{\scriptsize{3}}} and
        \normalsize{\textcircled{\scriptsize{4}}} correspond to Hubert-large, Clip-ViT-large, Baichuan-13B, Finetuned-Clip, and Finetuned-Baichuan-13B, respectively.}
        \label{tab:perfor}
        \begin{tabular}{|C{1.5cm}|C{2cm}|C{1.8cm}|C{1.8cm}|} 
            \hline
            \multirow{2}{*}{Model} & \multirow{2}{*}{Features} & \multicolumn{2}{c|}{Performance} \\ \cline{3-4}
            & & \textbf{Train\&Val\( \uparrow\)} & \textbf{MER-SEMI\( \uparrow\)} \\ 
            \hline
            \multirow{4}{1.5cm}{\centering BASELINE}
                & \normalsize{\textcircled{\scriptsize{0}}}+\normalsize{\textcircled{\scriptsize{1}}} & 78.77 & 84.34 \\ \cline{2-4}
                & \normalsize{\textcircled{\scriptsize{0}}}+\normalsize{\textcircled{\scriptsize{3}}} & 78.85 & 82.96 \\ \cline{2-4}
                & \normalsize{\textcircled{\scriptsize{0}}}+\normalsize{\textcircled{\scriptsize{1}}}+\normalsize{\textcircled{\scriptsize{2}}} & 79.34 & 86.86 \\ \cline{2-4}
                & \normalsize{\textcircled{\scriptsize{0}}}+\normalsize{\textcircled{\scriptsize{3}}}+\normalsize{\textcircled{\scriptsize{4}}} & 80.68 & 87.66 \\ 
            \hline
            \multirow{4}{1.5cm}{\centering AGT}
                & \normalsize{\textcircled{\scriptsize{0}}}+\normalsize{\textcircled{\scriptsize{1}}} & 79.12 & 83.43 \\ \cline{2-4}
                & \normalsize{\textcircled{\scriptsize{0}}}+\normalsize{\textcircled{\scriptsize{3}}} & 79.55 & 83.73 \\ \cline{2-4}
                & \normalsize{\textcircled{\scriptsize{0}}}+\normalsize{\textcircled{\scriptsize{1}}}+\normalsize{\textcircled{\scriptsize{2}}} & 82.89 & 87.51 \\ \cline{2-4}
                & \normalsize{\textcircled{\scriptsize{0}}}+\normalsize{\textcircled{\scriptsize{3}}}+\normalsize{\textcircled{\scriptsize{4}}} & \textbf{83.22} & \textbf{88.28} \\ 
            \hline
        \end{tabular}
    \end{threeparttable}
\end{table}


\begin{table}[tbp]
    \centering
    \fontsize{9}{12}\selectfont
    \caption{Results of different strategies. Please note that the audio modality consistently utilizes Hubert-large as the feature extractor. Meanwhile, the video and text modalities employ finetuned CLIP and finetuned Baichuan-13b as feature extractors respectively. `N' represents no additional strategy, `P' represents the use of pseudo labels, and `V' represents the use of the voting mechanism.}
    \label{tab:distortion_type}
    \begin{tabular}{|c|c|c|c|c|c|c|}
        \hline
        \multirow{2}{*}{Strategy} & \multicolumn{3}{c|}{BASELINE} & \multicolumn{3}{c|}{AGT} \\ \cline{2-7}
        & N & P & P + V & N & P & P + V \\ \hline
        \multirow{1}{*}{MER-SEMI} & 87.61 & 88.64 & 89.02 & 88.28 & 89.26 & \textbf{89.83} \\ \hline
    \end{tabular}
\end{table}

\bibliographystyle{ACM-Reference-Format}
\balance
\bibliography{authordraft}

\end{document}